\documentclass[runningheads]{llncs}
\usepackage[T1]{fontenc}
\usepackage{graphicx}
\usepackage{booktabs}
\usepackage{multirow}
\usepackage{amsmath}
\usepackage{amssymb}
\usepackage{dirtree}
\usepackage[subtle]{savetrees}
\usepackage{listings}
\usepackage{hyperref}
\usepackage{color}

\begin{document}
\title{A Simple Transformer Pipeline for Full-Key Side-Channel Attacks on Uncropped Datasets}
\titlerunning{Simple Transformer Architecture for Uncropped Full-Key Attacks}
%
\author{Jimmy Gammell \and Kaushik Roy}
\authorrunning{Jimmy Gammell and Kaushik Roy}
%
\institute{Elmore Family School of Electrical and Computer Engineering\\Purdue University, West Lafayette, IN 47907, USA\\\email{\{jgammell, kaushik\}@purdue.edu}}
\maketitle              
\begin{abstract}
Deep learning-based side-channel analysis has historically focused on single-byte targets and manually cropped traces, which risks discarding exploitable leakage. While recent work has proposed specialized architectures and resampling techniques to address this gap, the literature lacks a simple transformer baseline for simultaneous full-key attacks on uncropped traces. We present an open-source transformer implementation for uncropped full-key attacks which uses the standard transformer encoder backbone, adapting only the input and output layers to the side-channel setting. We release our implementation, training recipes, and pretrained weights for uncropped ASCADv1f, ASCADv1r, and CHES-CTF-2018 which achieve performance competitive with previously-reported results, while using less than 10GB of VRAM and requiring at most 3.34 hours of training on a single NVIDIA A6000 (\href{https://github.com/jimgammell/simple-transformer-pipeline-for-sca}{link}).
\end{abstract}

\section{Introduction}

Deep learning-based side-channel analysis (DLSCA) has emerged as a powerful tool for evaluating the vulnerability of cryptographic hardware to physical side-channel attacks \cite{picek2023sok}. However, much of the literature targets only one key byte, and crops traces to $\sim$1\% of their original length after manually identifying regions expected to leak that byte. This reduces modeling cost, but risks discarding exploitable leakage from unexpected instructions or locations \cite{COSADE:ESTLS22}. Such preprocessing is undesirable in defensive use cases where designers want to test the full execution for leaks without assuming where or why they occur.

Recent work has advanced beyond these simplified settings, successfully attacking multiple key bytes \cite{TCHES:PerWuPic22,COSADE:ESTLS22} and raw \cite{TCHES:LZCGL21,TCHES:BIKMPZ24}, long \cite{TCHES:HajChoMuk24}, or resampled \cite{TCHES:PerWuPic22} traces. In particular, \cite{TCHES:BIKMPZ24,TCHES:HajChoMuk24} have demonstrated the effectiveness of transformer-style architectures in long-trace settings. However, these specialized architectures depart significantly from standard transformer encoders, and the community lacks a simple baseline which leaves the standard transformer backbone unchanged. Additionally, no prior work to our knowledge has simultaneously attacked all key bytes from uncropped traces.

\begin{table}[]
    \centering
    \resizebox{\linewidth}{!}{\begin{tabular}{cccc}
    \toprule
    \textbf{Work} & \textbf{Trace preprocessing} & \textbf{Architecture} & \textbf{Target} \\
    \midrule
    EFSS~\cite{TCHES:PerWuPic22} & Downsampling/resampling & MLP/CNN & Full key (16 independent attacks) \\
    PART~\cite{TCHES:LZCGL21} & None (raw traces) & Specialized LSTM & Single byte \\
    EstraNet~\cite{TCHES:HajChoMuk24} & Cropping (long traces) & Specialized transformer & Single byte \\
    GPAM~\cite{TCHES:BIKMPZ24} & None (raw traces) & Specialized transformer & Single byte + random nonces \\
    Ours & None (raw traces) & Boilerplate transformer & Full key \\
    \bottomrule
\end{tabular}}
    \caption{Positioning of our artifact relative to representative prior work.}
    \label{tab:prior-work-positioning}
\end{table}

This artifact fills that gap with an installable PyTorch implementation of a standard transformer-based pipeline for full-key attacks on uncropped traces with no manual feature selection or cropping. We release training recipes and pretrained weights for the uncropped variants of ASCADv1f, ASCADv1r~\cite{JCEng:BPSCD20}, and CHES-CTF-2018~\cite{chesctf2018}, achieving competitive performance with results reported in prior work. Table~\ref{tab:prior-work-positioning} positions the artifact relative to representative prior work. Our aim is to reduce the engineering effort required for uncropped attacks and provide a deliberately simple baseline against which future side-channel-informed architectures can be benchmarked.

Our artifact supports OPTIMIST's goals by providing an open-source case study, reference implementation, and reproducible benchmark results for applying standard modern deep learning techniques to public side-channel datasets without manual feature selection or cropping. The intended users are DLSCA researchers studying uncropped attacks who want pretrained models without doing costly training and hyperparameter tuning, a simple and performant backbone to extend, or a reproducible baseline for boilerplate transformer-based attacks.

\section{Artifact overview}

\begin{figure}
\centering
\footnotesize
\dirtree{%
.1 config/ \DTcomment{YAML config files for dataset, model, and training configurations}.
.1 experiments/ \DTcomment{Experiment entrypoints and infrastructure}.
.1 src/uncropped\_transformers/ \DTcomment{Reusable importable library code}.
.2 datasets/ \DTcomment{Dataset loading and preprocessing}.
.2 models/ \DTcomment{Neural net architectures and building blocks}.
.2 training/ \DTcomment{Training loops, hyperparameter tuning}.
.2 evaluation/ \DTcomment{Accuracy, rank, and MTD metrics}.
}
\caption{High-level organization of our repository.}
\label{fig:repo-org}
\end{figure}

We release an open-source PyTorch repository for training, tuning, and evaluating profiled full-key attacks on uncropped side-channel datasets using transformers. This artifact can be installed alongside its dependencies by cloning our repository \href{https://anonymous.4open.science/r/optimist-submission-443E}{here} and typing \texttt{pip install -e .} inside its directory. To facilitate reuse, we separate importable library modules from experiment-specific entrypoints and infrastructure. See Fig.~\ref{fig:repo-org} for an overview of our repository's structure.

\paragraph{Architecture.} 

\begin{figure}[t]
    \centering
    \includegraphics[width=\linewidth]{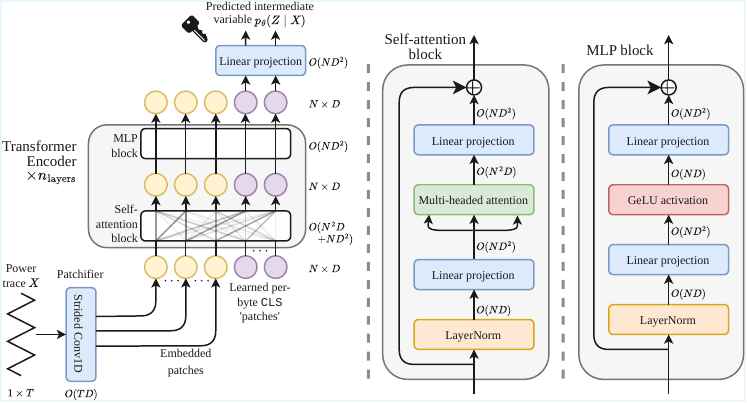}
    \caption{Overview of our full-key transformer pipeline. The input trace is divided into contiguous patches and linearly projected into token embeddings. One learned output token is appended for each targeted byte, and the resulting sequence is processed by a standard pre-norm transformer encoder. Each transformed output token is linearly projected to logits for its corresponding target byte.}
    \label{fig:transformer-diagram}
\end{figure}

The transformer is a widely-used sequence modeling architecture that can be applied to diverse domains and data modalities by representing inputs as token sequences while leaving the backbone largely unchanged. Accordingly, we use a standard pre-norm transformer encoder with rotary position embeddings and modify only the input and output layers for the side-channel setting. Inspired by vision transformers~\cite{dosovitskiy2021image}, we divide an input trace into contiguous patches and project them to the model's hidden dimension. We append one learned output token for each targeted byte, feed the resulting sequence through the transformer, and linearly project each transformed output token to logits for its corresponding byte. For all datasets, we use a single transformer to predict all 16 bytes of the first SubBytes output. For the ASCADv1 datasets we use the identity leakage model, and for CHES-CTF-2018 we use the multi-byte multi-bit model of \cite{EPRINT:WARPP23} because the identity model performed poorly in our experiments.

\paragraph{Training.} We train models using AdamW, applying weight decay to the weights (not biases) of the linear projections. The learning rate is linearly warmed up from zero to a base value over the first 5\% of training steps, then decayed using cosine annealing over the remaining steps. We optimize the mean cross-entropy loss across all target bytes, use a minibatch size of 256, and use gradient clipping with a maximum $\ell_2$ norm of 1. Each feature of input traces is standardized using its mean and standard deviation over the profiling set. We use a random 20\% of profiling traces for validation and use the remaining data for training. The attack set is treated as a test set and used only for final evaluation. We select the model checkpoint with the lowest single-trace correct-key rank on the validation set. Hyperparameter configurations and search spaces can be found in our repository.

\paragraph{Evaluation.} For all datasets we report the minimum number of attack traces required to correctly predict the key, denoted minimum traces to disclosure (MTD), averaged over 1k random permutations of the attack set. For ASCADv1 our model usually predicts the key from a single trace, so we additionally report the single-trace accuracy. An important consideration for full-key attacks is that per-byte performance does not uniquely determine full-key performance: if $A$ denotes the full-key accuracy and $A_i$ denotes the accuracy for byte $i,$ then $1 - \sum_i (1 - A_i) \leq A \leq \min_i A_i.$ For example, two bytes with 50\% accuracy may be jointly predicted with accuracy from 0 to 50\% depending on how correlated their errors are. Similarly, if $M$ is the full-key MTD and $M_i$ is the MTD for byte $i,$ then $M \geq \max_i M_i.$ Thus, we report both per-byte and full-key performance.

\section{Experimental results}

\begin{table}[t]
    \centering
    {\scriptsize\begin{tabular}{llll}
    \toprule
    \textbf{Dataset} &
    \textbf{Metric} &
    \textbf{Method} &
    \textbf{Result} \\
    \midrule

    \multirow{5}{*}{ASCADf~\cite{JCEng:BPSCD20}}
    & \multirow{3}{*}{\shortstack[l]{Byte 2--15 MTD $\downarrow$\\(best, mean, worst)}}
    & EFSS (CNN/ID)~\cite{TCHES:PerWuPic22}
    & \textbf{1}, $>$2.143k, $>$3k \\

    &
    & EFSS (MLP/ID)~\cite{TCHES:PerWuPic22}
    & 2, 5.571, 20 \\

    &
    & \textbf{Ours}
    & \textbf{1.000, 1.011, 1.055} \\

    \cmidrule(lr){2-4}

    &
    \multirow{2}{*}{Byte 2 MTD $\downarrow$}
    & EstraNet~\cite{TCHES:HajChoMuk24}
    & 13 \\

    &
    & \textbf{Ours}
    & \textbf{1.004} \\

    \midrule

    \multirow{8}{*}{ASCADr~\cite{JCEng:BPSCD20}}
    & \multirow{3}{*}{\shortstack[l]{Byte 2--15 MTD $\downarrow$\\(best, mean, worst)}}
    & EFSS (CNN/ID)~\cite{TCHES:PerWuPic22}
    & \textbf{1}, 4.643, 19 \\

    &
    & EFSS (MLP/ID)~\cite{TCHES:PerWuPic22}
    & \textbf{1}, 1.429, 3 \\

    &
    & \textbf{Ours}
    & \textbf{1.000, 1.001, 1.003} \\

    \cmidrule(lr){2-4}

    &
    \multirow{3}{*}{Byte 2 MTD $\downarrow$}
    & EstraNet~\cite{TCHES:HajChoMuk24}
    & 5 \\

    &
    & PART~\cite{TCHES:LZCGL21}
    & 8 \\

    &
    & \textbf{Ours}
    & \textbf{1.000} \\

    \cmidrule(lr){2-4}

    &
    \multirow{2}{*}{Byte 2 accuracy $\uparrow$}
    & GPAM~\cite{TCHES:BIKMPZ24}
    & 95.94\% \\

    &
    & \textbf{Ours}
    & \textbf{99.97\%} \\

    \midrule

    \multirow{3}{*}{CHES-CTF-2018~\cite{chesctf2018}}
    & \multirow{3}{*}{\shortstack[l]{Byte 0--15 MTD $\downarrow$\\(best, mean, worst)}}
    & EFSS (CNN/HW)~\cite{TCHES:PerWuPic22}
    & 317, $>$1.130k, $>$3k \\

    &
    & EFSS (MLP/HW)~\cite{TCHES:PerWuPic22}
    & \textbf{7}, 119.688, 288 \\

    &
    & \textbf{Ours}
    & 10.894, \textbf{11.384, 12.011} \\

    \bottomrule
\end{tabular}}
    \caption{Attack performance compared with results reported in prior work. Comparisons are not controlled for compute or hyperparameter tuning budget.}
    \label{tab:attack-performance}
\end{table}

\begin{table}[t]
    \centering
    {\scriptsize\begin{tabular}{lcccc}
\toprule
\textbf{Dataset} & \hspace{1em}\textbf{Parameters}\hspace{1em} & \hspace{1em}\textbf{FLOPs/step}\hspace{1em} & \hspace{1em}\textbf{Peak VRAM}\hspace{1em} & \hspace{1em}\textbf{Time/run} \\
\midrule
ASCADv1f~\cite{JCEng:BPSCD20} & 26.35M & 4.49T & 8.78GB & 3.34h \\
ASCADv1r~\cite{JCEng:BPSCD20} & 27.88M & 4.49T & 9.07GB & 2.80h \\
CHES-CTF-2018~\cite{chesctf2018} & 16.02M & 205G & 4.20GB & 0.36h \\
\bottomrule
\end{tabular}}
    \caption{Computational cost of training runs with recommended hyperparameters.}
    \label{tab:computational-cost}
\end{table}

We evaluate our pipeline on ASCADv1-fixed, ASCADv1-variable, and CHES-CTF-2018. Table~\ref{tab:attack-performance} compares our models with results reported in prior work under the corresponding targets and evaluation metrics. Our models achieve comparable or better performance on the reported metrics across all three datasets, establishing that the released models are competitive. In Table~\ref{tab:computational-cost} we report the parameter count, FLOPs, peak GPU memory, and wall-clock time of each training run under our recommended hyperparameters. Wall clock times are reported on a machine with an NVIDIA A6000 GPU, AMD Ryzen Threadripper PRO 5965WX CPU, and 128GB of RAM. Complete full-key and per-byte results, reproduction instructions, and additional computational scaling experiments can be found in the repository.

\section{Limitations}

Our experiments are limited to profiled attacks on first-order masked AES-128 implementations, targeting the 16 bytes of the first SubBytes output; performance in other settings remains to be evaluated. Comparisons with prior work are not matched for compute or hyperparameter tuning budget, so the strong performance of our trained models does not establish that the pipeline itself is superior. Observed performance differences may stem from architecture differences, more input features, more hyperparameter tuning, or regularization from jointly predicting multiple bytes. Finally, our experiments were conducted on machines with sufficient RAM for the OS to cache the full datasets. When traces are instead loaded repeatedly from a filesystem, data loading often becomes a bottleneck and substantially increases training time.

\bibliographystyle{splncs04}
\bibliography{abbrev3,crypto,biblio}

\end{document}